# Stretching of a Freely Jointed Chain in Two-Dimensions

Sara Iliafar,[†] Dmitri Vezenov,[#] and Anand Jagota[†$*]

[†]Department of Chemical Engineering and [$]Bioengineering Program, Lehigh University,

Bethlehem, Pennsylvania, PA 18015

[#]Department of Chemistry, Lehigh University, Bethlehem, Pennsylvania, PA 18015

*Corresponding Author: Anand Jagota (anj6@lehigh.edu)

## ABSTRACT

Although the stretching of polymers and biomolecules is important in numerous settings, their response when confined to two-dimensions is relatively poorly-studied. In this paper, we derive closed-form analytical expressions for the two-dimensional force-extension response of a freely-jointed chain under force control. Our principal results relate end-to-end distance to total force under two modes of stretching: i) when force is applied only to the free of the chain, and ii) when the applied force is distributed uniformly throughout the chain. We have verified both analytical models by Brownian dynamics simulation of molecules adsorbed strongly to a substrate. The total force required is always larger if distributed throughout the chain than when it is only applied to one end of the chain, and the nature of its divergence to infinity as extension approaches contour length is different.

1. INTRODUCTION

It is well known that the elastic and viscoelastic behavior of polymers derives from the force-displacement response of individual macromolecules.[1] Systems where elastic stretching of macromolecules occurs in three-dimensional (3D) settings are common and well-studied. Systems where polymer molecules are confined to a surface are less frequent and, perhaps for this reason, analytical results describing stretching in a two-dimensional (2D) case are not readily available in the literature.

For biological macromolecules such as DNA and polypeptides, this mechanical behavior of individual molecules plays an important physiological role. For this reason, numerous experimental studies have examined the 3D stretching of macromolecules via the use of atomic force microscopy and optical or magnetic tweezers,[2-4] electrophoretic stretching of DNA in uniform electric fields or flow,[5, 6] stretching of DNA under alternating current field,[7, 8] hydrodynamic focusing of multiple streams and the effect of velocity gradient created by hydrodynamic flow in contracting and expanding channels.[9, 10] To complement the experimental findings that assess the 3D stretching of a polymer, many theoretical models[11-15] and computer simulations have been explored by applying molecular dynamics and Monte Carlo approaches.[11, 12, 16, 17] Models range in complexity from the simple freely-jointed chain (FJC) and worm-like chain (WLC),[12, 16, 17] to all-atom representations in molecular dynamics.[18] Simpler models, such as the FJC and WLC, by providing explicit closed-form expressions relating force and stretch of the molecule, are particularly useful for interpretation and quantification of experiments. For example, it is well-known that the stretching of the freely jointed chain under force control is governed by the Langevin function,[16, 17] and a similar approximate force-extension relationship is available for the worm-like chain.[12] Similarly, exact relationships relating force required to peel an FJC or a WLC from a substrate can be found.[19-21] However, much less work has been conducted on either analytical or the numerical aspects of stretching in 2D.

The 2D stretching of polymers has been observed in systems involving the study of interfacial behavior of polymers at the air-water interface,[22] and combing of molecules via a meniscus alignment technique.[23, 24] In micro and nanofluidic systems, the transport of biopolymers such as DNA, RNA, and peptides has led to unprecedented advances in gene and restriction mapping.[25-29] The stretching of a (bio)polymer that is strongly adsorbed on a surface with one end fixed is often observed in systems involving separation of biomolecules via nanopillars and nanochannels.[30-32] Due to the increased surface area to volume ratio of nanofluidic environments, which increases the electrical and frictional effects, using nanostructures such as a capillary, nano channel, slit or a pore matrix (gel) results in a more efficient separation of macromolecules. Similarly, spontaneous trapping and 2D stretching behavior is also observed in nanochannels, where the choice of appropriate Debye length and the slit width (i.e. comparable to the radius of gyration of the DNA) results in entropic stretching of the macromolecule.[33-35] For example, in a study by Mailer, et al., the 2D stretching response of DNA to an external electric field was investigated by tethering one end of the molecule to the substrate and confining the entire molecule to the surface of a cationic lipid membrane.[36] While there has been a general lack of theoretical models for systems such as those described above, in a very recent study, Manca, et al., reported results for the stretching of a chain-like molecule due to a point and distributed forces both in 2D and 3D.[37, 38]

The principal results we derive in this paper are simple, closed-form, expressions relating force and stretch for a freely jointed chain confined to 2D (as well as some corresponding 3D results). Although several of our final expressions for stretching a molecule by a point or distributed force are similar to those discussed in Manca's recent work, we follow a more direct and transparent derivation. Furthermore, we provide approximate inverse expressions for the force-stretch relationship that are often needed, complement our analytical results by Brownian dynamics simulations, and validate one of the 2D results by comparison to experimental data. We expect that these results will be helpful to experimentalists for analyzing 2D stretching experiments.

## 2. METHODS

To verify our analytical results, we conducted Brownian Dynamics simulations of freely jointed chains in 3D and confined to a flat 2D surface, with and without self-avoidance. We used a program described previously elsewhere;[39] here we provide only a brief account. The freely jointed chain comprises N identical nodes connected by N-1 links. The vector form of the governing Brownian dynamics equation for bead n at position $r_n = (x_n, y_n, z_n)$, is written in terms of the viscous damping constant, $\xi_n$ (kg/s), a random force, $f_n^r$ (N), and the potential energy of the system as a function of coordinates of each link $E$:[40]

$$0 = -\xi_n \frac{d r_n}{dt} + f_n^r(t) - \nabla E_n(r_n) \tag{1}$$

The potential energy includes (i) possible repulsion between beads to model self-avoidance, (ii) attractive interaction with a surface to model adsorption, and (iii) constraints to enforce fixed bond length. One end was immobilized on a surface and force was applied to the other end either in a direction normal to the surface, to model 3D stretching, or in-plane, to model 2D stretching.

When modeling 2D stretching, we included an adsorption potential in the model that was sufficiently strong to ensure that all beads were strongly adsorbed to the surface and their motion confined to a frictionless surface. An adhesion free energy of 12 $k_B T$ per Kuhn length of the molecule was chosen based on our previous work representative of single-stranded DNA (ssDNA) as adequate for strong adsorbtion on a surface such as graphite.[19, 39]

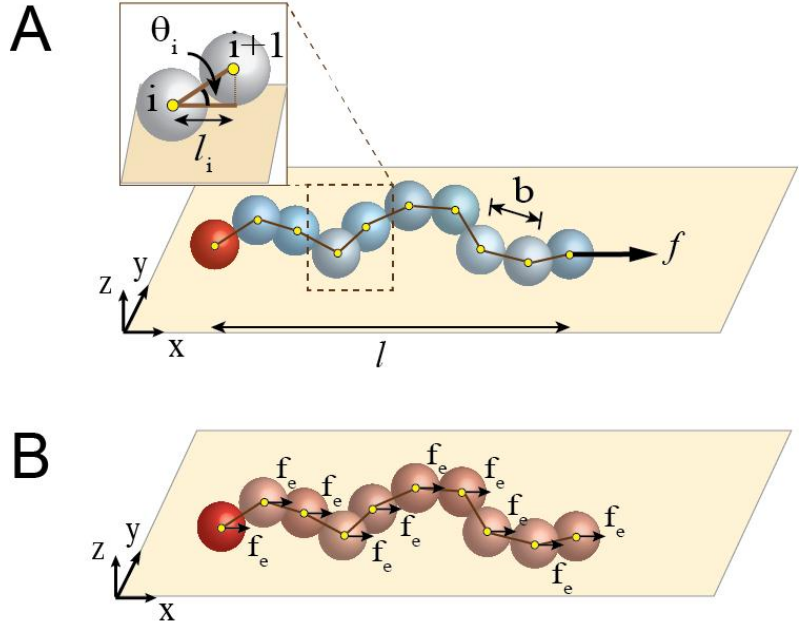

**Figure 1.** A schematic diagram of the freely jointed chain adsorbed on a solid substrate. The polymer chain is fixed to a point on the surface at one end (red node) and force is applied to the opposite end (A) or uniformly to each node (B). The molecule is represented by identical nodes, connected by freely jointed links.

## 3. RESULTS AND DISCUSSION

### 3.1 Force-displacement response of a freely jointed chain in 2D

The Langevin function relates average stretch of a freely jointed chain to applied force, f. This relationship is derived under fixed force and temperature constraints. A similar approach in 2D also yields closed form results. Consider the Helmholtz free energy of the externally loaded FJC:

$$A = U - TS - fl \qquad (2)$$

where $l = \sum_{i=1}^{N} l_i = \sum_{i=1}^{N} b\cos\theta_i$ is the end-to-end distance of the chain, with $b$ the Kuhn length, i.e., the distance between nodes (Figure 1). Combining Eq. (2) with the fundamental equation for energy:

$$dU = -pdV + TdS + fdl \tag{3}$$

we have

$$dA = -S\,dT - p\,dV - l\,df$$
$$l = -\left.\frac{\partial A}{\partial f}\right|_{T,V} = -\frac{\partial}{\partial f}\left(-k_B T \ln(Z_C)\right) \tag{4}$$

where $Z_C$ is the partition function for the FJC molecule. As done for the 3D FJC,[17] we need to consider only the conformational part of the partition function, since it is assumed that only orientation of the segments depends on force.

$$Z_C = \int \exp(fl/k_B T)d\mathbf{q} \tag{5}$$

where the integral is over all degrees of freedom that define the conformation of the molecule. Following Rubinstein and Colby,[17] $Z_C$ can be written as:

$$Z_{C,3D} = \left[\int_{\phi=0}^{2\pi}\int_{\theta=0}^{\pi} \exp(fb\cos\theta/k_B T)\frac{w_{3D}\sin\theta}{4\pi}d\theta\,d\phi\right]^N$$
$$= \left[w_{3D} \cdot \frac{\sinh(F)}{F}\right]^N \tag{6}$$

where $F = \frac{fb}{k_B T}$. Note the relatively minor departure from Rubinstein and Colby, who assume implicitly a density of states of one per steradian for the normalization constant.[19, 39] Using Eq. 4, one obtains the well-known stretch-force relationship for a freely jointed chain under a point force:

$$L_{pf,3D} = \frac{l}{Nb} = \coth(F) - \frac{1}{F} \equiv \mathcal{L}_{pf,3D}(F) \tag{7}$$

where, $\mathcal{L}_{pf,3D}(F)$ is the Langevin function.

Consider now the stretching of an FJC by a force applied to one end of the chain, i.e., point force (pf), while keeping the other end fixed and confining the entire chain to a 2D planar surface. The conformational partition function for the chain can be found by considering that in 2D each link, $i$, samples all orientations uniformly by angle, $\theta_i$, independently of all other links.

$$Z_{C,2D} = \int w_{2D} \exp\left(f\, l / k_B T\right) \prod_{i=1}^{N} d\theta_i \tag{8}$$

Here, the integral sign represents N integrals, one for each of the N links in the chain. Since the total length (projected on the axis of force) of the this FJC, $l = \sum_{i=1}^{N} l_i = \sum_{i=1}^{N} b \cos \theta_i$, Eq. 8 can be rewritten as:

$$\begin{aligned} Z_{C,2D} &= \int w_{2D} \exp\left(f \sum_{i=1}^{N} b \cos \theta_i / k_B T\right) \prod_{i=1}^{N} d\theta_i \\ &= \int \prod_{i=1}^{N} w_{2D} \exp\left(f\, b \cos \theta_i / k_B T\right) d\theta_i \end{aligned} \tag{9}$$

Because each integral is independent of the others:

$$Z_{C,2D} = \left[\int_{\theta=0}^{2\pi} w_{2D} \exp\left(f\, b \cos \theta / k_B T\right) d\theta\right]^{N} \tag{10}$$

The integral in Eq. 10 evaluates as[41]

$$\int_{\theta=0}^{2\pi} w_{2D} \exp\left(f\, b \cos \theta / k_B T\right) d\theta = 2\pi w_{2D}\, I_o(F), \tag{11}$$

where again $F = \dfrac{f\, b}{k_B T}$, and $I_0(F)$ is the modified Bessel functions of first kind of order '0',

resulting in:

$$Z_{C,2D} = \left[2\pi w_{2D}\, I_o(F)\right]^{N} \tag{12}$$

Using this expression for the partition function, the free energy of the FJC in 2D is:

$$A_{2D} = -k_B T \ln(Z_C) = -N k_B T \ln[2\pi w_{2D} I_o(F)] \tag{13}$$

Since we are under force-control, the end-to-end distance in the 2D case is

$$l = -\left.\frac{\partial A}{\partial f}\right|_{T,V} = -\frac{\partial}{\partial f}(-k_B T \ln(Z_C));$$

$$L_{pf,2D} = \frac{l}{Nb} = \frac{I_1(F)}{I_o(F)} \equiv \mathcal{L}_{pf,2D}(F) \tag{14}$$

where $I_1(F)$ and $I_0(F)$ are modified Bessel functions of first kind, and $\mathcal{L}_{pf,2D}(F)$ is defined as the 2D equivalent of the Langevin function under a point force.

For completeness, one can also list the known result for a one-dimensional (1D) case.[11] Although this case is not usually found in experimental situations, the stretching of a biomolecule such as ssDNA that is confined to a nanochannel may represent a situation that approximates this case.

$$L_{pf,1D} = \frac{l_{x,1D}}{Nb} = \tanh(F) \equiv \mathcal{L}_{pf,1D}(F) \tag{15}$$

In Figure 2A, we show that the force-displacement response in both the known 3D case (Eq. 7) and the 2D case (Eq. 14) matches the results of Brownian Dynamics simulations on a 21-bead FJC. Figure 2B shows good agreement between the simple analytical results, which neglect self-avoidance, and simulations that include a repulsive potential between beads to model self-avoidance. Not unexpectedly, it appears that the effect of self-avoidance is not significant for such short polymers. Also, as expected, stretching in 1D requires less force than in 2D; and 2D less than in 3D.[22]

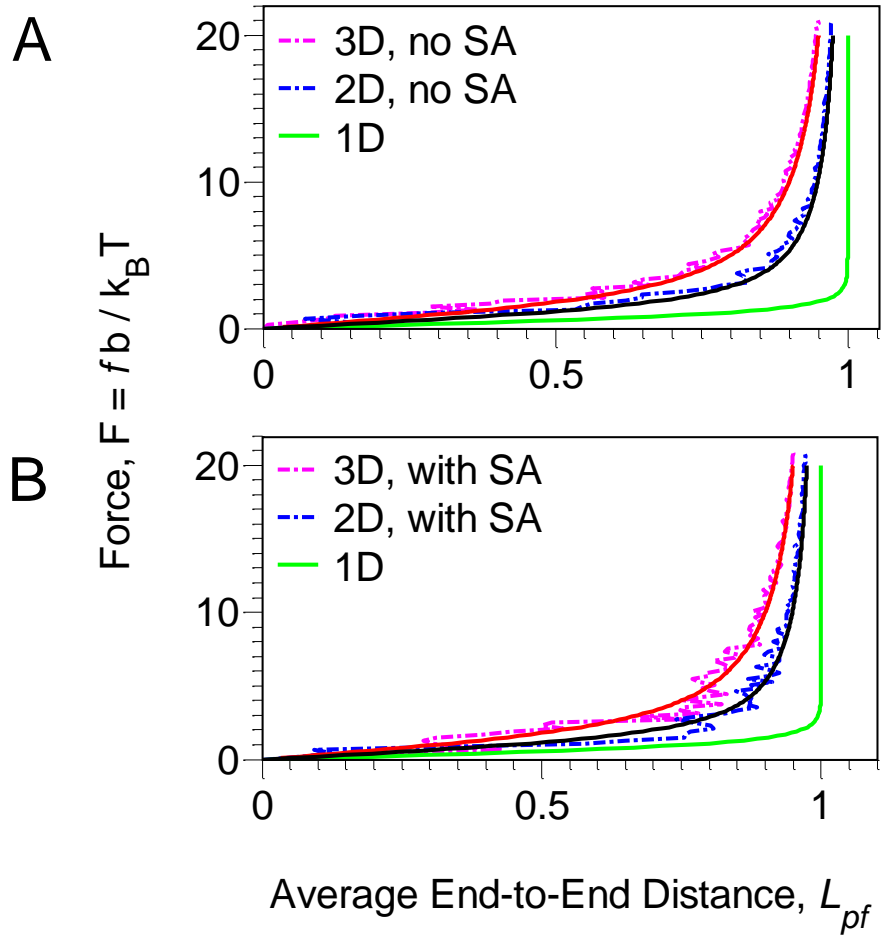

**Figure 2.** The force-displacement relationship for a freely jointed chain in 3D, 2D, and 1D under a point force (denoted as subscript pf in this figure). The solid curves represent the equilibrium results (equations 7, 14, and 15 for 3D, 2D, and 1D respectively), and the dashed lines are results obtained via Brownian dynamics simulations at slow-enough rates to be in equilibrium in A) the absence, and B) presence of self-avoidance .

### 3.2 Extension under an External Field

Stretching in 3D by force applied at the ends of the molecule is accomplished in numerous experiments that use force spectroscopy based on atomic force microscopy or optical/magnetic tweezers. In 2D, however, it is difficult to apply force only at the ends of the chain. Rather, most experiments employ a field, such as hydrodynamic flow or an electric field, that acts on all the beads. For example, an electric field applied to ssDNA tethered at one end will result in about the same force being applied to each charged (phosphate) group.[36] To help interpret such experiments, it is therefore useful to extend the results obtained above to the case where force is distributed along the molecule backbone.

Consider an FJC with *N* beads and *N-1* total links subjected to a field that provides a force $f_e$ on each bead. For example, if each bead carries a (net or effective) charge *q* and the molecule resides in an electric field *E*, then $f_e = qE$; if the molecule has N mobile beads, the total force is $f = N f_e$. Bead '1' is fixed but the remaining beads are free to move. Let $l_j$ be the distance between bead *j* and *j+1*, projected in the direction of field (force). Recall that, in the case where the external force was applied only to the last node, the projected location of only the last node was considered. In the case of distributed forces, the external force is applied equally to all nodes, and we must account for the projected location of all individual nodes. Let $L_i$ be the total distance from the fixed bead (bead '1') to bead 'i', i.e. $L_i = \sum_{j=1}^{i} l_j$ (Figure 1A). Then, the fundamental energy equation now has work contributions due to the movement of each of the charged beads in the electric field from the reference position (origin) to its final position $L_i$:

$$\begin{aligned} dU &= -pdV + TdS + \sum_{j=1}^{N} f_e dL_i \\ &= -pdV + TdS + f_e \sum_{i=1}^{N} dL_i \quad (16) \\ &= -pdV + TdS + f_e dl_{\Sigma} ; \qquad l_{\Sigma} \equiv \sum_{i=1}^{N} L_i \end{aligned}$$

Note that $l_{\Sigma}$ is the sum of the total projected lengths $L_i$'s from the fixed bead (bead '1') to all individual nodes 'i' and differs from the length, $l = \sum_{i=1}^{N} l_i = \sum_{i=1}^{N} b \cos \theta_i$, used in the first scenario, which was only the projected length from bead '1' to bead 'N'.  Following a Legendre transformation to switch to force control, we obtain:

$$dA = -S\,dT - p\,dV - l_\Sigma\,df_e$$

$$-\left.\frac{\partial A}{\partial f_e}\right|_{T,V} = l_\Sigma \tag{17}$$

To calculate the free energy A, we first need the conformational partition function, for which we note that the argument of the Boltzmann factor now has contributions from each of the mobile beads.  Therefore, the conformational partition function is

$$Z_{C,2D} = \int w_{2D}\,\exp\left(\sum_{i=1}^{N}\frac{f_e L_i}{k_B T}\right)\prod_{j=1}^{N}d\theta_j \tag{18}$$

The argument of the exponential can be written as

$$\sum_{i=1}^{N}\frac{f_e L_i}{k_B T} = \left(\frac{f_e L_1 + f_e L_2 + f_e L_3 + \ldots}{k_B T}\right) = \frac{1}{k_B T}\begin{pmatrix} f_e b\cos\theta_1 + \\ f_e b[\cos\theta_1 + \cos\theta_2] + \\ f_e b[\cos\theta_1 + \cos\theta_2 + \cos\theta_3] + \\ \ldots\ldots \end{pmatrix}$$

$$= \frac{1}{k_B T}\begin{pmatrix} N f_e b\cos\theta_1 + \\ (N-1)f_e b\cos\theta_2 + \\ (N-2)f_e b\cos\theta_3 + \\ \ldots\ldots + f_e b\cos\theta_N \end{pmatrix} = \sum_{i=1}^{N}\frac{(N-i+1)f_e b\cos\theta_i}{k_B T} \tag{19}$$

Substituting (19) into (18)

$$Z_{C,2D} = \int w_{2D}\,\exp\left(\sum_{i=1}^{N}\frac{(N-i+1)f_e b\cos\theta_i}{k_B T}\right)\prod_{j=1}^{N}d\theta_j$$

$$= \prod_{i=1}^{N}w_{2D}\int_{0}^{2\pi}\exp\left(\frac{(N-i+1)f_e b\cos\theta_i}{k_B T}\right)d\theta_i \tag{20}$$

and, considering that all angles are independent, this equation can be rewritten as:

$$Z_{C,2D} = \prod_{i=1}^{N} w_{2D} \int_{0}^{2\pi} \exp\left(\frac{i f_e b \cos\theta_i}{k_B T}\right) d\theta_i \tag{21}$$

By applying the identity of the modified Bessel functions of the first kind of order v (used previously in Eq. 11),[41] $I_\nu(z) = \frac{1}{2\pi} \int_0^{2\pi} \exp(z \cos\theta) \cdot \cos(\nu \theta) d\theta$ with $z = \frac{i f_e b}{k_B T}$, the previous function becomes:

$$Z_{C,2D} = (2\pi w_{2D})^N \prod_{i=1}^{N} I_o(iF_e) \tag{22}$$

$$F_e = f_e b / k_B T \tag{23}$$

Using Eq. 17, we obtain for $l_\Sigma$:

$$A = -k_B T \ln(Z_{2D}) = -k_B T \left[ N \ln(2\pi w_{2D}) + \sum_{i=1}^{N} \ln(I_o(iF_e)) \right] \tag{24}$$

$$l_\Sigma = \sum_{i=1}^{N} L_i = \sum_{i=1}^{N} \left( \sum_{j=1}^{i} l_j \right) = -\left.\frac{\partial A}{\partial f_e}\right|_{T,V} = \sum_{i=1}^{N} ib \frac{I_1(iF_e)}{I_o(iF_e)} \tag{25}$$

We can interpret $l_\Sigma$ as:

$$\begin{aligned} l_\Sigma &= \sum_{i=1}^{N} L_i = L_N + L_{N-1} + \ldots + L_2 + L_1 = \\ &= (l_1 + l_2 + \ldots + l_{N-1} + l_N) + (l_1 + l_2 + \ldots + l_{N-1}) + \ldots + (l_1 + l_2) + l_1 = \\ &= l_N + 2l_{N-1} + \ldots + (N-1)l_2 + N l_1 = \sum_{j=1}^{N} (N+1-j) l_j \end{aligned} \tag{26}$$

Comparing Eq. 25 and 26, we observe that:

$$l_j = b \frac{I_1([N+1-j]F_e)}{I_o([N+1-j]F_e)} \tag{27}$$

The physical interpretation of Eq. 27 can be given as follows. Although an equal magnitude of an external force, $f_e$, is applied to each bead, the effective force acting on each beads is determined by how far along the FJC each node is located at with respect to the direction of the force applied. In other words, while $f_e$ is applied to the last bead 'N', twice the magnitude of $f_e$ is applied to bead 'N -1', bead 'N-2' experiences three times the force $f_e$, and so on. The effective applied force on each bead 'j' (i.e. N = j) is obtained by evaluating Eqs. 10 through 14 and substituting the corresponding multiples of $f_e$ for F, so that:

$$l_N = b\frac{I_1(F_e)}{I_o(F_e)}; l_{N-1} = b\frac{I_1(2F_e)}{I_o(2F_e)} \quad .... \quad l_1 = b\frac{I_1(NF_e)}{I_o(NF_e)} \tag{28}$$

Of particular interest is the end-to-end distance of the chain, $L_N$, which is the projected distance from bead '1' to bead 'N' given by

$$\begin{aligned} l_{x,2D} \equiv L_N = (l_1 + l_2 + ....l_N) = \sum_{i=1}^{N} b\frac{I_1(iF_e)}{I_o(iF_e)} \\ L_{ff,2D} = \frac{l_{x,2D}}{Nb} = \frac{1}{N}\sum_{i=1}^{N}\frac{I_1(iF_e)}{I_o(iF_e)} \end{aligned} \tag{29}$$

To compare the case where the force is distributed along the chain ($f_e$ applied to each node) with the case where it is applied only to the end (force $f$), let the total force in each case be the same, i.e. set $f_e = f/N$. In Figure 3, we show the force-displacement relationship obtained from Eqs. 14 and 29 compared to corresponding Brownian Dynamics simulations. As might be expected, for the same magnitude of the total force applied, the molecule will always be more extended when force is applied only to one end of the chain.

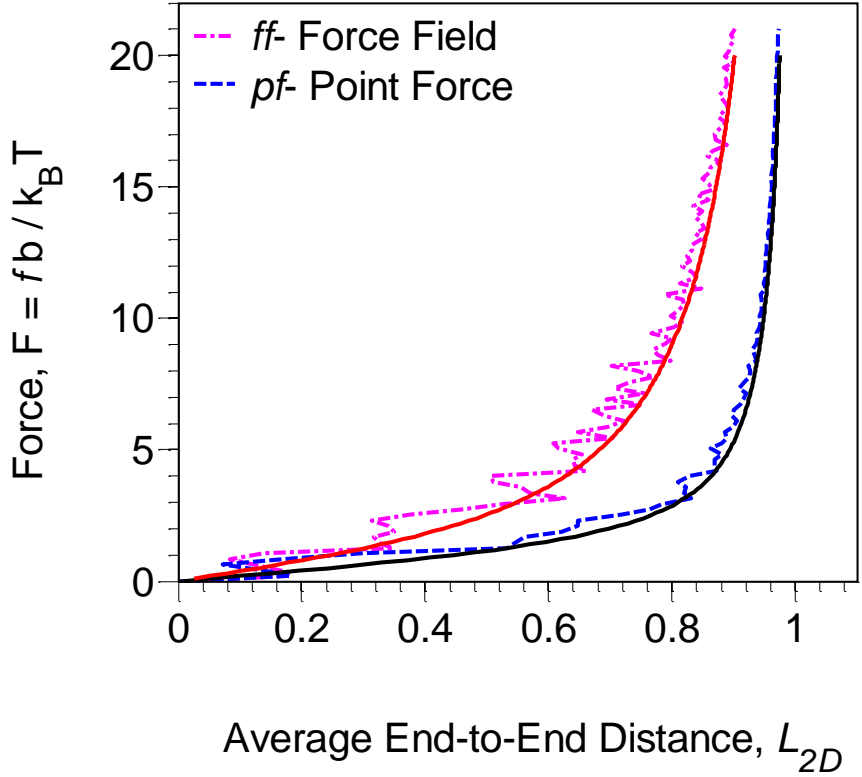

**Figure 3.** The force-displacement relationship for a non-self-avoiding freely jointed chain presenting the elastic response of the chain when it is stretched in 2D due to i) force applied to one end of the chain i.e. point force (dashed blue line), and ii) a force field applied to the entire chain (dashed magenta line). The stretching behavior of the chain is correctly predicted by $L_{pf,2D} = \dfrac{I_1(F)}{I_o(F)}$ and $L_{ff,2D} = \dfrac{1}{N}\sum_{i=1}^{N}\dfrac{I_1(iF_e)}{I_o(iF_e)}$ as shown by the solid black and the solid red line, correspondingly.

To assess this stretching behavior in the limit of long chains, we rewrite the end-to-end distance for a chain under a force field as:

$$l_{x,2D} \equiv L_N = b\sum_{i=1}^{N}\frac{I_1(iF_e)}{I_o(iF_e)} = b\int_{i=1}^{i=N}\frac{I_1(iF_e)}{I_o(iF_e)}di = \left.\frac{b\ln(I_o(iF_e))}{F_e}\right|_{i=1}^{i=N} \quad (30)$$

We have replaced the summation by an integral, since the force is distributed in very small quanta $(F_e = F/N)$ as the number of Kuhn segments, N, increases to a large value. Therefore, the end-to-end distance of the chain becomes:

$$l_{x,2D} = \frac{b \ln \left(I_o\left(N F_e\right)\right)}{F_e} = \frac{N b \ln \left(I_o(F)\right)}{F}$$

or  (31)

$$L_{ff,2D} = \frac{l_{x,2D}}{N b} = \frac{\ln \left(I_o(F)\right)}{F} = \mathcal{L}_{ff,2D}(F)$$

which is a simpler result for distributed force than the one in Eq. 29, since it removes the summation. Here, $\mathcal{L}_{ff,2D}(F)$ is the equivalent of the Langevin function for the 2D case under a force field. This extension can be compared with the end-to-end distance of a long chain, to which force is applied only at one end:

$$\frac{L_{ff,2D}(F_e = F/N)}{L_{pf,2D}(F)} = \frac{\frac{\ln \left(I_o(F)\right)}{F}}{\frac{I_1(F)}{I_o(F)}} \quad (32)$$

To test our analytical result for stretching a FJC under an electric field, we used a data set previously published by Mailer, et al. for elastic response of λ-DNA to external fields in two dimensions.[36] In this experiment, the authors confined the DNA to a cationic lipid membrane, and tethered one end of the molecule to a bead immobilized on the surface. They then applied external electric field of various strengths, and measured the projected extension of the molecule. We extracted numerical force-field data from Figure 3 of Mailer's paper, and plotted against our theoretical results. Note that the experimental data were reported as the extension of the biopolymer, $l_{x,2D} = L_{ff,2D} N b$ (μm) vs. electric field, E (V/cm). The molecule used in this experiment is double stranded DNA (dsDNA) with contour length, $L_c = N b$, of 20 ± 1 μm (48502 base pairs, where each base pair has elongation length of 0.44 nm).[36] Typically, dsDNA is modeled as a worm-like chain, however, by taking the Kuhn length, b, of this chain to be twice its persistence length ($l_p$ = 65 nm[36]), one can treat the λ-DNA as a freely jointed chain. The applied force, $f = N f_e$ is defined in terms of electric field, E, as follows:

$$f = L_c q E \tag{33}$$

where, $q$ is the effective electrophoretic line charge density and reported to be $0.6 \pm 0.1$ e per Kuhn length in this experiment.[36] The experimentally reported electric field strength values were converted into applied force, which was then non-dimensionalized as $F = \dfrac{f b}{k_B T}$. Figure 4 compares this experimental set of results to results for a FJC both under a force field (eq. 31) as well as a point force at one end (eq. 14), with no adjustable parameters. It is apparent that the result for distributed force, eq. 31, matches the experimental behavior of this long chain.

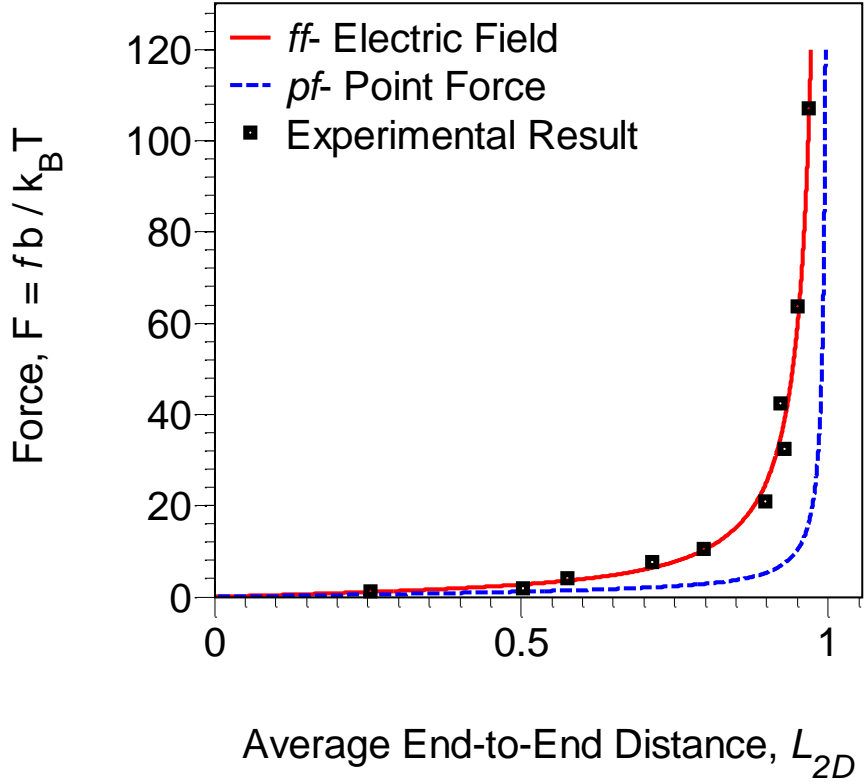

**Figure 4.** The elastic response of a long freely jointed chain when it is stretched in 2D due to i) a force field applied to the entire chain e.g. under an electric field (solid line) predicted by $L_{ff,2D} = \dfrac{\ln(I_o(F))}{F}$, and ii) force applied to one end of the chain, i.e. point force (dashed line) predicted by $L_{pf,2D} = \dfrac{I_1(F)}{I_o(F)}$. The experimental data[36] for stretching λ-DNA confined in 2D and under an electric field closely follow the predicted results for distributed force mode of stretching.

The stretching of molecules due to a force field is often observed in gel electrophoresis and translocation of molecules through pores. In such applications, the molecule takes a 3D conformational form. For this reason, we used a similar approach as shown for the 2D case above, to derive an expression for the 3D stretching of a FJC under a force field. Starting with the conformational partition function,

$$Z_{C,3D} = \int_{\phi=0}^{\phi=2\pi} \int_{\theta=0}^{\theta=\pi} \exp\left(\sum_{i=1}^{N} \frac{f_e L_i}{k_B T}\right) \prod_{i=1}^{N} \frac{w_{3D}}{4\pi} \sin\theta_i \, d\theta_i \, d\phi_i \qquad (34)$$

where, using equation (19), $\sum_{i=1}^{N} \frac{f_e L_i}{k_B T} = \sum_{i=1}^{N} \frac{(N-i+1) f_e b \cos \theta_i}{k_B T} = \sum_{i=1}^{N} \frac{i f_e b \cos \theta_i}{k_B T}$, Eq. 34

becomes:

$$Z_{C,3D} = -(w_{3D})^N \prod_{i=1}^{N} \int_{\theta=0}^{\theta=\pi} \exp\left(\frac{i f_e b \cos \theta_i}{k_B T}\right) \frac{d(\cos \theta_i)}{2}$$
$$Z_{C,3D} = (w_{3D})^N \prod_{i=1}^{N} \left(\frac{\sinh(iF_e)}{iF_e}\right) \qquad F_e = \frac{f_e b}{k_B T} \tag{35}$$

As before, the free energy of the FJC in 3D is written as:

$$A = -k_B T \ln(Z_{C,3D}) = -k_B T N \ln(w_{3D}) - k_B T \sum_{i=1}^{i=N} \ln\left(\frac{\sinh(iF_e)}{iF_e}\right) \tag{36}$$

Using Eq. 17, the sum of the total projected lengths, $l_\Sigma \equiv \sum_{i=1}^{N} L_i = -\left.\frac{\partial A}{\partial f_e}\right|_{T,V}$, becomes:

$$l_\Sigma = ib \sum_{i=1}^{N} \left\{\coth(iF_e) - \frac{1}{iF_e}\right\} \tag{37}$$

Although the sum of the total projected lengths is obtainable from computer simulations, it would be difficult to compare the results shown in Eq. 37 to experimental results. We can, however, experimentally measure the end-to-end extension of the molecule and, therefore, need an expression that correctly defines the projected end-to-end length of the chain as a function of applied force. We used the same interpretation as described earlier for the 2D stretching under a force field, and found that the projected length of the chain becomes:

$$L_{ff,3D} = \frac{l_{x,3D}}{Nb} = \frac{1}{N} \sum_{i=1}^{N} \left\{\coth(iF_e) - \frac{1}{iF_e}\right\} \tag{38}$$

For very long molecules, Eq. 38 is rewritten as:

$$L_{ff,3D} = \frac{l_{x,3D}}{Nb} = \frac{1}{N} \int_{i=1}^{N} \left\{ \coth(iF_e) - \frac{1}{iF_e} \right\} di = \frac{1}{NF_e} \ln\left( \frac{\sinh(NF_e)}{N \cdot \sinh(F_e)} \right)$$

$$L_{ff,3D} = \frac{1}{F} \ln\left( \frac{\sinh(F)}{N \cdot \sinh(F/N)} \right) \tag{39}$$

For large molecules, $\sinh(F/N) \to F/N$, and Eq. 39 becomes:

$$L_{ff,3D} = \frac{1}{F} \ln\left( \frac{\sinh(F)}{F} \right) = \mathcal{L}_{ff,3D}(F) \tag{40}$$

where, $\mathcal{L}_{ff,3D}(F)$ is the equivalent of the Langevin function for the 3D case under a force field. Comparing this extension to the end-to-end distance of a long chain under a point force, we find that:

$$\frac{L_{ff,3D}(F_e = F/N)}{L_{pf,3D}(F)} = \frac{\ln\left( \frac{\sinh(F)}{F} \right)}{F \coth(F) - 1} \tag{41}$$

Figure 5 shows this comparison of predicted stretch for point vs. distributed load for both the 2D and the 3D cases. In both cases, for small forces (when extension is proportional to force), end-to-end deflection for distributed force is half of that for a point force applied at the terminus of the molecule. As expected, when large enough force is applied to the chain, so that it is fully stretched out to its contour length, $L_c$, this extension approaches unity.

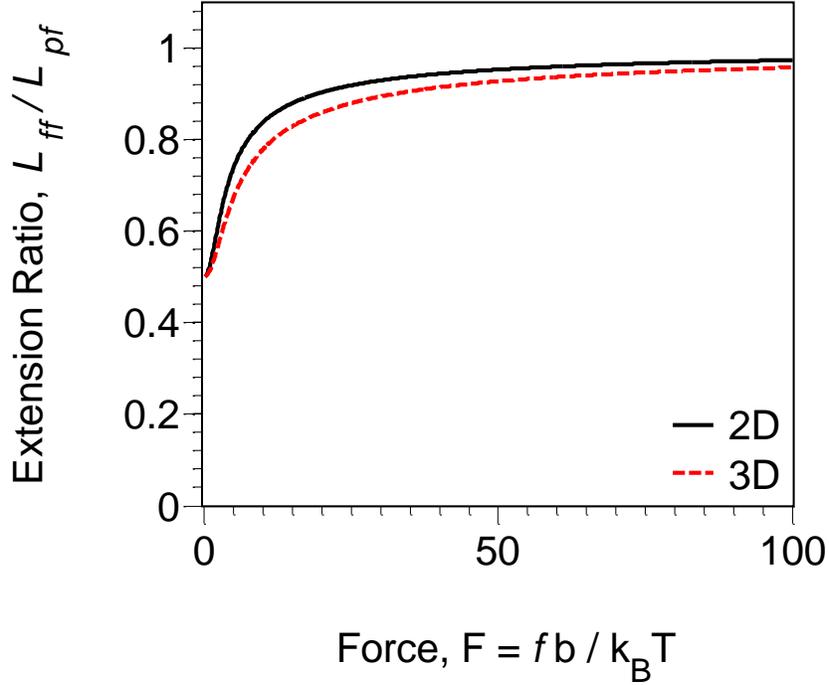

**Figure 5.** The elastic response of a long 3D freely jointed chain when a force field is applied to the entire chain vs. when the force is applied to one end of the chain. This result shows that for short extensions, twice as much force is required in both the 2D and the 3D case to stretch the chain under an electric field than when force is applied only to one end of the chain.

The expressions that we have reported so far for the stretching of a chain are explicit in terms of force; however, it is often useful to know the force explicitly in terms of the normalized extension, $L = \dfrac{l_x}{Nb}$. An approximation for the Inverse Langevin function for point force stretching in 3D has been reported in the literature,[42]

$$F_{pf,3D} = \mathcal{L}^{-1}_{3D}(L_{pf,3D}) = L_{pf,3D} \cdot \left( \dfrac{3 - L^2_{pf,3D}}{1 - L^2_{pf,3D}} \right) \quad (42)$$

To derive a similar approximation as the inverse langevin function for 3D stretching under a point force, consider first the stretching behavior of the chain at the limits of small and large extensions. For small arguments in the 3D case under a point force, Eq. (7) can be approximated

by using $\coth(F) \approx \frac{1}{F} + \frac{F}{3}$, and

$$F_{pf,3D} = 3 L_{pf,3D} \qquad \text{for } F \ll 1 \qquad (43)$$

Similarly, for small arguments in the 2D case under a point force (Eq. 14), $I_1(F) \approx F/2$; $I_o(F) \approx 1$.

Therefore, $\mathcal{L}_{2D}(F) \approx F/2$, and the dimensionless force, $F = \frac{fb}{k_B T}$, becomes:

$$F_{pf,2D} = 2 L_{pf,2D} \qquad \text{for } F \ll 1 \qquad (44)$$

and, in the limit of low force, the 1D stretching relationship is approximated as:

$$F_{pf,1D} = L_{pf,1D} \qquad \text{for } F \ll 1 \qquad (45)$$

Comparing Eq. 43 through 45, it becomes apparent that for small lengths, the force-displacement relationship is governed by the system's dimensionality, as is expected for Gaussian chains.[43] As shown in Figure 5, for small force, the force-extension relationship under a force field is twice as large as that under a point force. Therefore, for small arguments in the 3D (Eq. 40) and 2D (Eq. 31) cases under a force field, the stretching is approximated as following, correspondingly:

$$F_{ff,3D} = 6 L_{ff,3D} \qquad \text{for } F \ll 1 \qquad (46)$$

$$F_{ff,2D} = 4 L_{ff,2D} \qquad \text{for } F \ll 1 \qquad (47)$$

Using the asymptotic expression for the modified Bessel's functions of order $v$ for large arguments:[41]

$$I_v(z) \approx \frac{e^z}{\sqrt{2\pi z}} \left(1 + \frac{(1-2v) \cdot (1+2v)}{8z}\right) \qquad (48)$$

we find the response for the 2D case under a point force (Eq.14) for large forces to be:

$$L_{pf,2D}(F) = \frac{I_1(F)}{I_o(F)} \approx \frac{8F-3}{8F+1} \qquad \text{for } F \gg 1$$

or  (49)

$$F_{pf,2D} = \frac{(3+L_{pf,2D})}{8(1-L_{pf,2D})} \qquad \text{for } F \gg 1$$

A good approximation for the inverse 2D function that satisfies both limits (Eqs. 44 and 49) is:

$$F_{pf,2D} = \mathcal{L}^{-1}_{pf,2D}(L_{pf,2D}) \approx \frac{L_{pf,2D}(8 - L_{pf,2D} - 3L^2_{pf,2D})}{4(1-L^2_{pf,2D})} \qquad (50)$$

However, the distributed force results, equations (31,40), do not have large-force limits that can be written as simple rational functions. Using eq. (48) in (40) shows that, in the limit of large force,

$$L_{ff,3D} = \frac{1}{F}\ln\left(\frac{\sinh(F)}{F}\right) \to \frac{1}{F}\ln\left(\frac{\exp(F)}{F}\right) = 1 - \ln(F)/F;$$
$$1 - L \approx \ln(F)/F; \qquad (51)$$
$$F \approx \frac{\ln(1/(1-L))}{(1-L)}$$

Using eq. (48) in (31) shows that the force in the 2D, distributed force, case also diverges as

$F \sim \dfrac{\ln(1/(1-L))}{(1-L)}$ as $L$ approaches unity. However, this converges slowly to the exact result, and is

not very useful as a guide to obtain an approximate inverse function that satisfies the limits of both small and large forces. Therefore, we adopt empirically the general form of Eq. 42:

$$F = \frac{aL(b-L^2)}{(1-L^2)} \qquad . \qquad (52)$$

To ensure that the force-extension behavior of the chain for small arguments is satisfied, we set the limit of Eq. 51 for small arguments, i.e. $F \approx abL$, equal to those limits given by Eq. 46 and 47. By fitting eq. (52) to the exact result in the range ($F<50$), we find values of $a$ and $b$.

Figure 6 shows the exact and approximate invers force-distance relationships obtained from both the Langevin functions and their inverse functions in 2D and 3D. We have collected all the results in Table 1 along with the maximum error of the inverse functions.

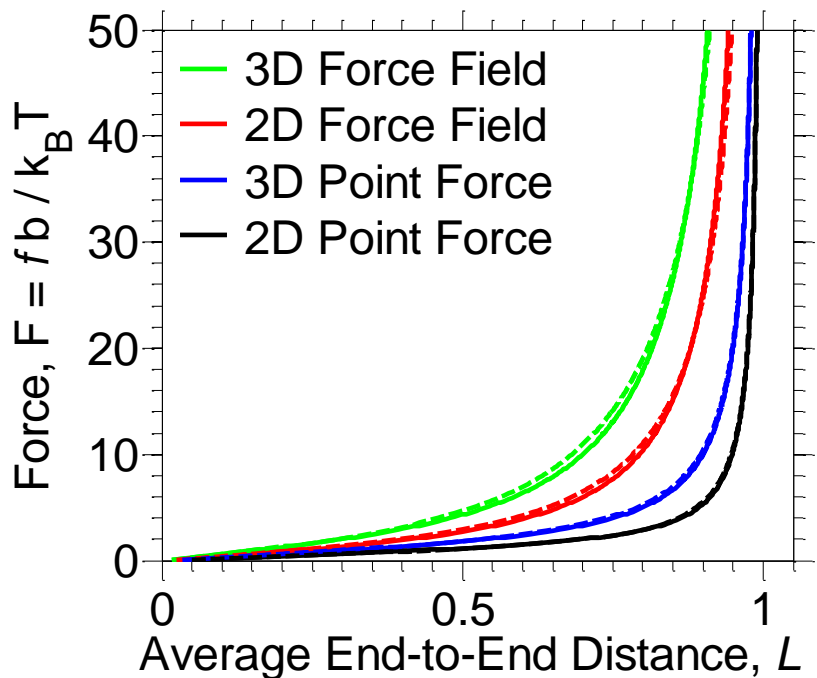

**Figure 6.** The force-displacement relationship for 3D and 2D elastic responses of a freely jointed chain presented along with approximate inverse functions. The solid curves represent the exact results; the dashed lines show the approximate inverse functions.

|  |  | 2D |  | 3D |  |
|---|---|---|---|---|---|
| **Point Force** | $L(F)$ | $L = \dfrac{I_1(F)}{I_o(F)}$ | (Eq. 14) | $L = \coth(F) - \dfrac{1}{F}$ | (Eq. 7) |
|  | $F(L)$ | $F \approx \dfrac{L(8 - L - 3L^2)}{4(1-L^2)}$ | (Eq. 50) | $F \approx L \cdot \left(\dfrac{3-L^2}{1-L^2}\right)$ | (Eq. 42) |
|  |  | $\varepsilon = 4.3\%$ |  | $\varepsilon = 4.9\%$ |  |
|  |  | $F = 2L$ for $L \to 0$ |  | $F = 3L$ for $L \to 0$ |  |
|  |  | $F \approx \dfrac{(3+L)}{8(1-L)}$ for $L \to 1$ |  | $F \approx \dfrac{1}{1-L}$ for $L \to 1$ |  |
| **Force Field** | $L(F)$ | $L = \dfrac{\ln(I_o(F))}{F}$ | (Eq. 31) | $L = \dfrac{1}{F}\ln\left(\dfrac{\sinh(F)}{F}\right)$ | (Eq. 40) |
|  | $F(L)$ | $F \approx \dfrac{3L(8/3 + L^2)}{2(1-L^2)}$ | (Eq. 52) | $F \approx \dfrac{4L(3/2 + L^2)}{(1-L^2)}$ | (Eq. 52) |
|  |  | $\varepsilon = 9.9\%$ |  | $\varepsilon = 9.9\%$ |  |
|  |  | $F = 4L$ for $L \to 0$ |  | $F = 6L$ for $L \to 0$ |  |

**Table 1.** Compilation of extension-force relations for a freely jointed chain in 2D & 3D, and under point or distributed force. We also present approximate inverse functions for force in terms of extension along with the maximum error, $\varepsilon$, in the range $F<50$.

## 4. CONCLUSIONS

The 2D stretching of polymers and biomolecules often occurs during processes such as molecular combing (or meniscus alignment) and in separation techniques that use nanofluidic systems. In some cases, as in molecular combing, the applied force acts on the chain at a single point at the air-water interface. On the contrary, when molecules are transported and separated in nanofluidic systems, often an electric field is applied across the sample. In this case, the total force is distributed over the entire molecule. To aid the quantitative analysis of such experiments, in this paper, we have presented analytical expressions to describe the 2D and 3D stretching of a freely jointed chain under two modes of stretching: i) when force is applied only to one end of the chain, and ii) when the applied force is distributed uniformly throughout the chain. We have provided expressions that describe the force-extension relationship as force as a function of extension, as well as extension explicitly in terms of force. These results were verified using Brownian dynamics simulations and, in the case of stretching in 2D by distributed force, compared to experimental results.

## ACKNOWLEDGEMENT

This work was supported by the National Science Foundation through grant CMMI-1014960. We would like to acknowledge useful discussion with Prof. Philip Blythe.